%% file: main.tex
\documentclass[runningheads]{llncs}
\usepackage[T1]{fontenc}
\usepackage{graphicx}
\usepackage{xcolor}
\usepackage{amsmath}
\usepackage{listings}
\usepackage{enumitem} 
\usepackage{amsthm}

\usepackage{algorithm}
\usepackage{algpseudocode}
\usepackage{rotating} 

\usepackage{comment}

\usepackage{multirow}
\usepackage{pgf}
\usepackage{tikz}
\usepackage[clock]{ifsym}
\usepackage[edges]{forest}
\usetikzlibrary{angles}
\usetikzlibrary{positioning}

\usetikzlibrary{arrows,shapes,automata,backgrounds,petri}
\usetikzlibrary{angles,shadows.blur,positioning,backgrounds}

\usetikzlibrary{quotes,graphs}

\usepackage{featuremodel}

\newcommand{\titan}{\textsc{Titan}}

\newcommand{\tessla}{TeSSLa}

\newcommand{\hlola}{HLola}
\newcommand{\code}[1]{\textsf{#1}}

\begin{document}
%
\title{Efficient Construction of Reachability Graphs for Petri Net Product Lines}
%
%
\author{Elena G\'{o}mez-Mart\'{i}nez\inst{1}\orcidID{0000-0002-7753-3345} \and Jos\'{e} Ignacio Requeno Jarabo\inst{1}\orcidID{0000-0001-5111-8357}}
\authorrunning{E. G\'{o}mez-Mart\'{i}nez and J.\ I. Requeno Jarabo}
\institute{Departamento de Sistemas Inform\'{a}ticos y Computaci\'{o}n, \\
Universidad Complutense de Madrid\\
Calle del Prof.\ Jos\'{e} Garc\'{i}a Santesmases, 9, 28040 Madrid, Spain\\
\email{mariaelena.gomez@ucm.es, jrequeno@ucm.es}}
\maketitle              
\begin{abstract}
This paper presents a set of algorithms for computing the reachability graph of Petri Net Product Lines (PNPLs). These algorithms address the combined challenges of concurrency and variability that arise from product-line configurations. The proposed approach integrates symbolic state representations with family-based variability handling to generate a compact, parameterised reachability graph that captures behaviour across all products without exhaustive product enumeration. 
The main contributions are threefold. First, we introduce a symbolic state encoding adapted to PNPL semantics. Second, we define a family-preserving successor generation procedure that applies feature constraints during exploration. Third, we propose reduction techniques to mitigate state-space explosion, including on-the-fly merging of equivalent symbolic states and selective abstraction of irrelevant state details. We prove soundness and completeness of the construction with respect to standard per-product semantics and analyse computational complexity. 
An implementation integrated into our modelling tool demonstrates substantial savings in memory and time compared with naive product-based exploration, while preserving diagnostic and verification capabilities. The results indicate that the method enables practical reachability analysis for realistically sized product-line models, thereby facilitating verification and design-space exploration in configurable concurrent systems.

\end{abstract}
%
%
%
\input{introduction}
\input{background}
\input{algorithms}

\input{reachability_graph}
\input{tool_support}

\input{related}

\input{conclusions}

\section*{Acknowledgments}
This work has been supported by the Spanish MINECO/FEDER projects AwESOMe (PID2021-122215NB-C31) and the Region of Madrid project DESAFíO-CM (TEC-2024/COM-235).

%
%
%
\bibliographystyle{splncs04}
\bibliography{references}

\end{document}

%% file: introduction.tex
\section{Introduction}
\label{sec:introduction}

Petri nets (PNs) constitute a well-established graph-based modelling formalism for concurrent systems~\cite{Petri62,Murata89}. 
Their graphical notation and firm mathematical foundations have made them a de facto choice for analysing control-flow, resource usage and behavioural properties in distributed and industrial settings.

Product-line approaches enable the compact specification of many closely related system variants by capturing variability explicitly. 
A Petri Net Product Line (PNPL)~\cite{Gomez-MartinezL19,Gomez-MartinezLG21} combines both ideas: it encodes a family of Petri nets together with presence conditions and feature constraints, avoiding duplication and enabling analyses that operate at the product-line level. 
Previous work has shown that structural analyses (e.g., checking net classes, structural soundness or free-choice properties) can be ``lifted'' to PNPLs, yielding significant efficiency gains compared to per-product analysis~\cite{Gomez-MartinezLG21}.

Despite these advantages, important gaps remain when attempting to reuse the rich set of dynamic analyses developed for traditional Petri nets in the PNPL setting. 
A central example is the reachability graph: a fundamental artefact for verification (e.g., model checking, liveness, deadlock detection), performance reasoning and diagnostic tasks in classical PN theory~\cite{Murata89}. 
For PNPLs, however, the definition and construction of a family-preserving reachability graph are not yet standardised nor widely implemented. 
The interaction between variability (presence conditions, feature constraints) and state-space exploration introduces several challenges: naive per-product enumeration is infeasible for realistic product counts, symbolic encodings must preserve feature information to avoid loss of per-product semantics, and state-space explosion is exacerbated by the combinatorial interplay of control and variability~\cite{Gomez-MartinezLG21}.

These limitations hinder the applicability of PNPLs in scenarios that require behavioural analysis across variants, such as global safety checks, variant-aware verification or automated debugging of configurable concurrent systems. 
Bridging this gap requires: (i) a precise semantics for PNPL states that combines net marking with variability information, (ii) successor-generation algorithms that enforce feature constraints during exploration, and (iii) reduction and merging strategies that keep the family-level state space tractable while preserving soundness and completeness with respect to per-product semantics.

This paper addresses those challenges. 
We present a formalisation and a set of algorithms to construct a family-based reachability graph for PNPLs. 
Our approach defines a symbolic state representation that pairs markings with presence conditions, a family-preserving successor relation, and state-space reduction techniques to mitigate explosion. 
We prove soundness and completeness of the construction with respect to the standard per-product semantics, provide complexity considerations, and report an implementation integrated in the \titan{} modelling environment that leverages existing tools for Petri net analysis~\cite{JensenKW07}. 
Experimental results on benchmark PNPL models illustrate practical gains over naive product enumeration and demonstrate the utility of the reachability graph for verification and diagnosis across product families.

The rest of the paper is organised as follows.
After this introduction, Section~\ref{sec:background} presents the essential background for this work.
Section~\ref{sec:algorithms} details the algorithms for successor generation and state reduction.
Section~\ref{sec:tool_support} shows the architecture of the \titan{} framework that implements our solution.
Section~\ref{sec:related} positions our contribution with respect to prior work on PNPLs and Petri net analysis.
Finally, Section~\ref{sec:conclusions} summarises the conclusions and outlines future work.

%% file: background.tex
\section{Background}
\label{sec:background}
In this section, we introduce essential concepts about Petri nets (Section~\ref{sec:PN}), and PNPLs (Section~\ref{sec:pnpl}).

\subsection{Petri Nets}
\label{sec:PN}
A Petri net (PN) is a graphical and mathematical modelling tool used to describe concurrent systems \cite{Murata89}. Petri nets provide a natural and effective means to represent logical interactions among system components or activities, including synchronisation, sequencing, concurrency, and conflict resolution. 

A PN model comprises two key elements: a net structure, which represents the static aspects of the system as a weighted, bipartite directed graph (places and transitions); and a marking, which denotes the system state by means of tokens distributed across the net structure~\cite{Murata89}. The graphical representation of places, transitions and tokens employs circles, rectangles and dots, correspondingly. 

The evolution of markings through transition firings is known as the \emph{token game} and simulates the behaviour of the system \cite{JensenKW07}. The reachability graph (RG) of a PN is the directed graph whose vertices represent all markings reachable from the initial marking, with edges labelled by transitions corresponding to firings between markings. The RG is a central artefact to analyse dynamic properties such as reachability, boundedness, liveness, reversibility, and deadlock freedom \cite{Murata89}. However, RG construction suffers from state explosion for larger systems, motivating more efficient techniques such as simulation-based analysis \cite{zurawski1994petri} or model checking \cite{girault2013petri}.

\subsection{Petri Net Product Lines}
\label{sec:pnpl}


Petri Net Product Lines (PNPLs)~\cite{Gomez-MartinezL19,Gomez-MartinezLG21,Gomez-MartinezG23} integrate Petri nets with Software Product Line (SPL) engineering principles~\cite{Northrop,SPLs} to represent a collection of similar systems in a unified and compact form. A PNPL thus denotes a family of Petri nets that can be derived through distinct admissible configurations. 

The adoption of a single artefact enables the concurrent analysis of all potential derivations, thereby eliminating the need for independent evaluations of each variant. Thus, this approach enables modelling and analysis of all system variants simultaneously, avoiding redundant effort for each configuration.

\begin{definition}[Petri Net Product Line~\cite{Gomez-MartinezL19}] 
A Petri Net Product Line is a tuple $\mathit{PNPL} = (\mathit{PN}, FM, \Phi)$ where:
\begin{itemize}
  \item $\mathit{PN}$ is the 150\% Petri net, which includes all places, transitions, and arcs across the product family;
  \item $FM$ is a feature model that defines the variability space; and
  \item $\Phi$ is a set of presence conditions, expressed as logical formulas, that specify under which feature combinations each element appears in a product variant.
\end{itemize} 
Elements whose presence conditions evaluate to \emph{false} are removed from the 150\% net to derive the configured net for a specific variant. Variability affects only the structure of the PN; the marking, which represents the dynamic state, remains unchanged.
\end{definition}

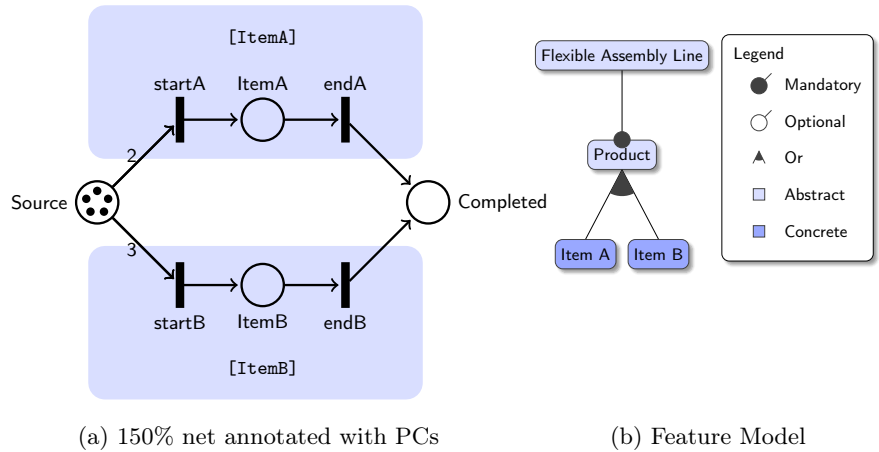
\begin{figure}
    \centering
    \begin{tabular}{p{7cm}c}
        \multirow{2}{*}{\input{150mm}} & \\
            & \input{fm} \\
        \\
        \\
        \\
        \\
        \\
        \multicolumn{1}{c}{(a) 150\% net annotated with PCs} & (b) Feature Model\\
    \end{tabular}
    \caption{A PNPL representing a Simplistic Flexible Assembly Line.}
    \label{fig:running}
\end{figure}

As illustrated in Figure~\ref{fig:running}, the model represents a very simple, Flexible Assembly Line capable of producing two different types of products, \code{ItemA} and \code{ItemB}. \code{ItemA} requires two tokens for its assembly, and three for \code{ItemB}. This PNPL has 3 variants: \code{ItemA}, \code{ItemB} and \code{(ItemA} $\wedge$ \code{ItemB}). This running example will be used in Section~\ref{sec:algorithms} to illustrate the construction and pruning of the feature-annotated reachability graph.

Structural properties --- such as marked graphs, state machines, free-choice, and extended free-choice structures --- are formally analysed using lifted analysis techniques~\cite{SalayFRSC14}. These techniques translate structural properties into first-order logical propositions that can be efficiently resolved through satisfiability (SAT) solvers~\cite{Gomez-MartinezLG21}. Furthermore, by transforming a PNPL into a constraint programming formulation~\cite{Brown2006}, P- and T-invariants can be examined at the product-line level through a lifted incidence matrix representation, as described in~\cite{Gomez-MartinezG23}.

Prior to this work, it was not possible to construct the RG of a PNPL in order to analyse behavioural properties. 


%% file: 150mm.tex
\scalebox{1.1}{
\begin{tikzpicture}[
on grid,node distance = 5cm and 5cm,thick,scale=0.8, 
every node/.style={scale=0.75},
every transition/.style={fill,minimum width=7mm,minimum height=1mm},
stransition/.style={transition,fill,minimum height=7mm,minimum width=1mm} ,
font=\sffamily]
    \node[place,tokens=5, label = left:Source](S){};
    \node[place,tokens=0, right = 4 of S, label = right:Completed](C){};
    
    \node[right = 1 of S](d1){};
    \node[above = 1 of S](d2){};
    \node[below = 1 of S](d3){};

    \node[place,tokens=0, right = 2 of d2, label = above:ItemA](PA){};
    \node[place,tokens=0, below = 2 of PA, label = below:ItemB](PB){};
    
    \node[stransition, left = 1 of PA, label = above:startA](sa){}
        edge[pre] node[left]{2} (S)
        edge[pre] (S) 
        edge[post] (PA);
    \node[stransition, right = 1 of PA, label = above:endA](ea){}
        edge[pre] (PA)
        edge[post] (C);

    \node[stransition, left = 1 of PB, label = below:startB](sb){}
        edge[pre] node[left]{3} (S)
        edge[pre] (S)
        edge[post] (PB);
    \node[stransition, right = 1 of PB, label = below:endB](eb){}
        edge[pre] (PB)
        edge[post] (C);

    \node (label1) [above= 1 of PA] {\texttt{\textbf{[ItemA]}}};
    \node (label2) [below= 1 of PB] {\texttt{\textbf{[ItemB]}}};
        
    \begin{pgfonlayer}{background}
        \filldraw [line width=4mm,join=round,blue!85!cyan!15]
        (label1.north -| d2.east) rectangle (ea.south -| C.west)
        (sb.north -| d3.east) rectangle (label2.south -| C.west)
        ;
    \end{pgfonlayer}
\end{tikzpicture} 
}

%% file: fm.tex
\scalebox{0.7}{
\begin{forest}
  disjunction tree,
  [Flexible Assembly Line 
    [Product,abstract,mandatory, or={1/l}
        [Item A]
        [Item B]
    ]
  ]
\end{forest}
}

%% file: algorithms.tex
\section{Construction of the Reachability Graph of a PNPL}
\label{sec:algorithms}

This section outlines the formal procedure for transforming a PNPL into a reachability graph (RG) augmented with feature annotations, named the feature-annotated RG (fRG). 

\input{fRG}
\input{PNPL_2_fRG}

%% file: fRG.tex
\subsection{Feature-Annotated Reachability Graph}
\label{sec:frg}

The Reachability Graph (RG) of a Petri Net is a state-space representation that enumerates all possible markings reachable from the initial marking through the firing of transitions. Each node in the net represents a reachable marking, while each directed edge corresponds to a transition firing that causes the system to evolve from one marking to another. 

In a PNPL, variability is introduced by associating transitions, places, or arcs with presence conditions over features that govern their inclusion under specific product configurations. In order to facilitate dynamic property verification -- for example, to ensure liveness or safety – across all potential products, a unified behavioural model capturing this variability must be generated. 

The following formal definition introduces the concept of the feature-annotated reachability graph (fRG).

\begin{definition}[Feature-Annotated Reachability Graph]\\
Let $\mathit{PNPL} = (FM, PN, \Phi)$ be a Petri Net Product Line, where $PN = (P, T, F, W, M_0)$ is the 150\% Petri net with its initial marking, and $FM = (F, C)$ is the feature model with feature set $F$ and constraints $C$. The feature-annotated reachability graph (fRG) of a PNPL is a tuple
\[
\mathit{fRG} = (V, E, f)
\]
where:
\begin{itemize}
  \item $V \subseteq \mathcal{B}(P)$ is the set of reachable markings from $M_0$, with $\mathcal{B}(P)$ denoting the set of all possible markings over $P$;
  \item $E \subseteq V \times T \times V$ is the set of directed edges, where each $(M, t, M') \in E$ iff $M \xrightarrow{t} M'$ is a valid firing in the 150\% net $PN$; and
  \item $f : E \to 2^{F}$ is a labelling function that assigns to each edge $e = (M, t, M') \in E$ a presence condition $f(e) \subseteq F$ (or its corresponding feature formula), where $F$ is the set of features defined in $FM$.
\end{itemize}
\end{definition}

Extending the RG concept to the PNPL setting requires integrating variability information. The fRG therefore associates each transition firing with a feature expression (presence condition) that governs its activation across different product configurations. This enriched state space enables a unified analysis of behavioural properties across the entire product line.

%% file: PNPL_2_fRG.tex
\subsection{From PNPL to Feature-Annotated RG}
\label{sec:pnpl2frg}

When constructing the fRG from a PNPL model, special care is required to ensure correct state generation, particularly in scenarios with multiple tokens and overlapping transitions. A single marking $M$ may enable multiple transitions associated with distinct feature expressions; if these transitions fire using different tokens, the successor marking $M'$ may implicitly combine mutually exclusive features, yielding semantically invalid states. 

\subsubsection{Conflict Detection Filter.}
To prevent such invalid states, we introduce a \emph{conflict detection filter} that systematically excludes markings derived from incompatible feature combinations. 

Formally, for any path $M \to M'$, if $M'$ results from firing transitions with feature expressions $\varphi_1, \varphi_2, \ldots$ such that $\bigwedge_i \varphi_i \not\models \top$ under the feature model constraints $C$, then $M'$ is pruned from the state space. The filter operates as follows:

\begin{enumerate}
  \item \textbf{Feature Path Tracking}: Each marking $M \in V$ maintains a feature path $\Phi_M = \bigwedge_{i=1}^k \varphi_i$, recording the cumulative conjunction of feature expressions of all transitions fired from $M_0$ to reach $M$.
  \item \textbf{Compatibility Check}: Before adding an edge $(M \xrightarrow{t} M')$ with feature expression $\varphi_t$, check that $\Phi_M \land \varphi_t \models \top$ under $C$.
  \item \textbf{Pruning Rule}: Prune $M'$ if $\Phi_{M'} = \Phi_M \land \varphi_t$ is unsatisfiable or violates the feature model constraints (e.g.\ XOR, \emph{requires}, or \emph{excludes} relationships). Formally:
    \[
      M' \notin V \quad \text{if} \quad \Phi_{M'} \not\models \top \;\lor\; \Phi_{M'} \not\models C.
    \]
  \item \textbf{Token-Level Validation}: For markings with multiple tokens in input places, ensure that the tokens enabling $t$ originate from compatible feature branches. If some $p \in \bullet t$ contains tokens constrained by $\varphi_p^1$ and $\varphi_p^2$ with $\varphi_p^1 \land \varphi_p^2 = \bot$, firing $t$ is disabled.
  \item \textbf{Incremental Update}: Compute $\Phi_{M'} = \Phi_M \land \varphi_t$ incrementally to enable efficient SAT solving or constraint propagation.
\end{enumerate}

This strategy guarantees that every state in the fRG corresponds to a valid product configuration, eliminating false positives from mixed-feature interactions while preserving completeness for valid feature combinations.

\subsubsection{Construction Procedure.}
The fRG, integrating the conflict detection filter, is constructed through the following recursive procedure:
\begin{enumerate}
  \item Initialise with the 150\% net and its initial marking $M_0$.
  \item For each enabled transition $t \in T$ in marking $M$, compute the successor $M'$ according to the standard firing rule.
  \item Apply the conflict detection filter by checking $\Phi_M \land \varphi_t \models \top$.
  \item If the check succeeds, annotate the edge $(M \xrightarrow{t} M')$ with the presence condition $\chi = \Phi_M \land \varphi_t$ and add $M'$ to the state space.
  \item Recursively process all valid markings reachable from $M_0$.
\end{enumerate}

\input{algorithm_w_conflict}

%% file: algorithm_w_conflict.tex
\begin{algorithm}[h!]
\caption{Algorithm for generating an fRG from its PNPL (with conflict detection)}
\label{alg:PNPL2fRG}
\begin{algorithmic}[1]
    \Require $P$, $T$, $A$, $M_0$, $fs$ ($fs \subseteq FM$), feature model constraints $C$
    \Ensure $fRG \leftarrow$ A set of feature-annotated paths
    \State $V \leftarrow \emptyset$ \Comment{Set of visited markings}
    \State $Q \leftarrow \emptyset$ \Comment{Processing queue}
    \State $fRG \leftarrow \emptyset$ \Comment{Feature-annotated reachability graph}
    \State $\Phi \leftarrow \emptyset$ \Comment{Map: marking $\mapsto$ feature path}
    \State $\Phi[M_0] \leftarrow \top$
    \State $Q.\textit{add}(M_0)$
    
    \While{$Q \neq \emptyset$}
        \State $M \leftarrow Q.\textit{pop}()$
        \For{each $t_i$ in $T$}
            \State $ia \leftarrow t_i.\textit{getInputArcs}(A)$
            \State $oa \leftarrow t_i.\textit{getOutputArcs}(A)$
            \State $ip \leftarrow ia.\textit{getPlaces}(P)$ 
            \State $op \leftarrow oa.\textit{getPlaces}(P)$
            
            \If{$M[t_i.\textit{enabled}(fs)]$} \Comment{Token + local feature enablement}
                \State $M' \leftarrow M$
                \State $M'.\textit{decreaseTokens}(ip, ia.\textit{getWeights}())$
                \State $M'.\textit{increaseTokens}(op, oa.\textit{getWeights}())$
                \If{$\forall p \in ip : M[p] \geq ia[p].\textit{weight}$} \Comment{Firing precondition}
                    \State $\varphi_t \leftarrow t_i.\textit{getFeatureFormula}()$
                    \State $\Phi_{\textit{cand}} \leftarrow \Phi[M] \land \varphi_t$
                    \If{$\textit{SAT}(\Phi_{\textit{cand}} \land C)$} \Comment{Conflict detection filter}
                        \State $fRG.\textit{add}(\text{Edge}(M \xrightarrow[\varphi_t]{} M'))$
                        \If{$M' \notin V$}
                            \State $V.\textit{add}(M')$
                            \State $\Phi[M'] \leftarrow \Phi_{\textit{cand}}$
                            \State $Q.\textit{add}(M')$
                        \EndIf
                    \EndIf
                \EndIf
            \EndIf    
        \EndFor
    \EndWhile
    \State \Return $fRG$
\end{algorithmic}
\end{algorithm}

Algorithm~\ref{alg:PNPL2fRG} implements the construction of the feature-annotated reachability graph (fRG) from a PNPL by means of a breadth-first exploration of the state space. It takes as input the set of places $P$, transitions $T$, arcs $A$, the initial marking $M_0$, the feature set $fs \subseteq FM$, and the feature model constraints $C$, and produces the corresponding feature-annotated reachability graph $fRG$ together with feature paths $\Phi$ for each reachable marking.

The algorithm first initialises four data structures (lines~1--4): the set of visited markings $V$, the processing queue $Q$, the (initially empty) reachability graph $fRG$, and the map $\Phi$ that records, for each marking $M$, the cumulative feature path $\Phi_M$. The initial feature path is set to $\Phi[M_0] = \top$ (line~5). The exploration is seeded by inserting the initial marking $M_0$ into the queue $Q$ (line~6). The main loop (lines~7--32) then performs a standard breadth-first search: at each iteration, a marking $M$ is dequeued from $Q$ (line~8), and all transitions $t_i \in T$ are considered for firing (line~9).

For each transition $t_i$, the algorithm retrieves its input and output arcs (lines~10--11), and the corresponding input and output places (lines~12--13). It then checks whether $t_i$ is enabled under the current feature configuration $fs$ and marking $M$ (line~14), using the predicate $M[t_i.\textit{enabled}(fs)]$, which encapsulates both the usual token-enabling condition and the satisfaction of the local presence condition for $t_i$ with respect to $fs$.

If $t_i$ is enabled, a successor marking $M'$ is computed (lines~15--17). This is done by first cloning the current marking $M$ into $M'$ (line~15), then decreasing the number of tokens in all input places according to the weights of the input arcs (line~16), and finally increasing the tokens in all output places according to the weights of the output arcs (line~17). The explicit firing precondition $\forall p \in ip : M[p] \geq ia[p].\textit{weight}$ is then checked (line~18) to ensure that the transition is indeed fireable from $M$ with respect to the underlying Petri net semantics.

If the firing is valid at the token level, the algorithm applies the conflict detection filter (lines~19--21). It retrieves the feature formula $\varphi_t$ associated with $t_i$ (line~19), computes the candidate feature path $\Phi_{\textit{cand}} = \Phi[M] \land \varphi_t$ (line~20), and checks whether $\Phi_{\textit{cand}} \land C$ is satisfiable using the feature model constraints (line~21). Only if this satisfiability check succeeds does the algorithm add a feature-annotated edge to $fRG$ (line~22). This edge connects $M$ to $M'$ and is labelled with $\varphi_t$, thus realising the construction of the fRG where each transition is annotated with its presence condition. To avoid re-processing the same marking multiple times, the algorithm only enqueues $M'$ if it has not been visited before (lines~23--26). In that case, $M'$ is added to the visited set $V$ (line~24), its feature path is stored as $\Phi[M'] \leftarrow \Phi_{\textit{cand}}$ (line~25), and it is pushed into the queue $Q$ for further exploration (line~26). The loop continues until $Q$ becomes empty (line~32), meaning that all reachable feature-consistent markings have been explored. 

Finally, the completed feature-annotated reachability graph $fRG$ is returned (line~33), guaranteeing termination and completeness with respect to the reachable, feature-consistent state space of the configured PNPL.

\subsubsection*{Computational Complexity.}
Let $|V|$ denote the number of reachable markings and $|T|$ the number of transitions in the 150\% net. Algorithm~\ref{alg:PNPL2fRG} performs a breadth-first exploration of the state space: each marking is dequeued from $Q$ and processed at most once (lines~7--9), and for each processed marking $M$ all transitions $t_i \in T$ are inspected (line~9). The inner operations for a given pair $(M,t_i)$---retrieving arcs and places (lines~10--13), checking enablement (line~14), computing the successor marking $M'$ (lines~15--17), applying the firing precondition and conflict detection filter (lines~18--21), and testing membership in $V$ (lines~23--26)---all run in constant time assuming bounded in/out degree, bounded or amortised cost for checking $\Phi_{\textit{cand}} \land C$, and efficient data structures for sets and queues. Hence, the overall time complexity is
\[
  O(|V| \cdot |T|),
\]
linear in the product of the number of reachable markings and transitions.

The algorithm avoids fully unfolding the PNPL for each individual feature configuration and instead constructs a symbolic RG in which variability is encoded through transition annotations. Nevertheless, if the number of reachable markings grows exponentially with the number of tokens or places, the construction may still suffer from state explosion in practice.

%% file: reachability_graph.tex
\subsubsection{Illustrative Example.}
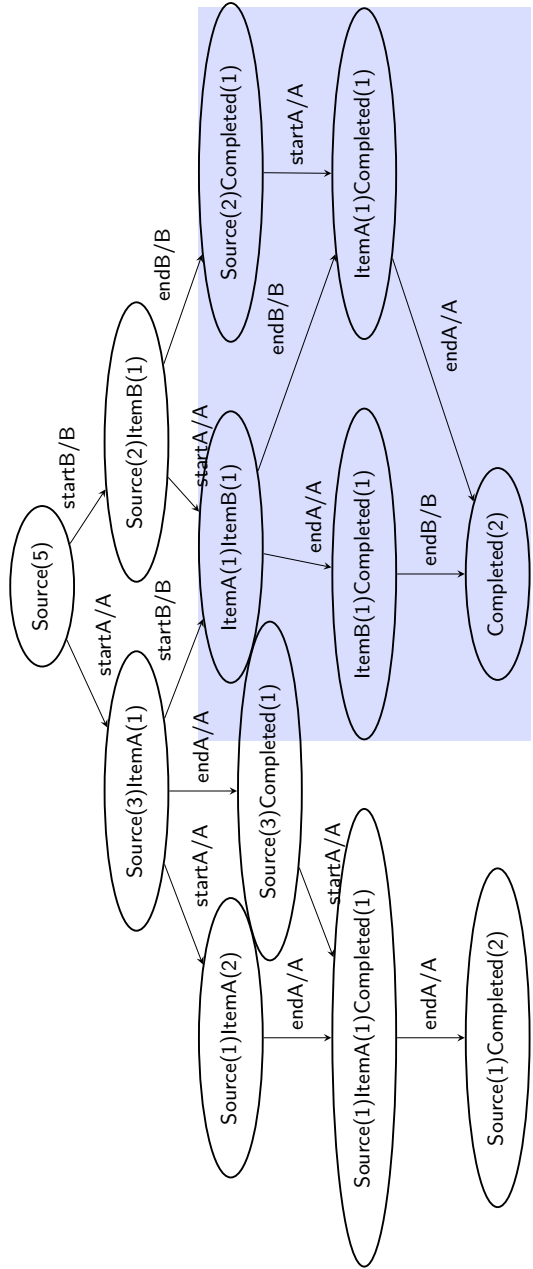
\begin{figure}[h!]
     \centering
     \begin{sideways}
        \begin{minipage}{17cm}
            \input{pruned_rg}
        \end{minipage}
    \end{sideways}
    \caption{Feature-annotated reachability graph for PNPL representing a Simplistic Flexible Assembly Line.}
    \label{fig:pruned_rg}
\end{figure}

Figure~\ref{fig:pruned_rg} shows the reachability graph for PNPL in Figure~\ref{fig:running}.
Nodes represent the marking of the places, and edges represent the firing transitions. 
For instance, \code{Source(3)ItemA(1)} means that place \code{ItemA} receives one token, and \code{Source} retains three tokens when transition \code{startA} from product line \code{[ItemA]} is fired.

Nevertheless, some configurations that are naively depicted by this reachability graph are actually not feasible because of the feature constraints: e.g., reaching \code{ItemA(1)ItemB(1)} from the previous state when firing \code{startB} transition (PNPL cannot arbitrarily reach states of product \code{[ItemB]} from product \code{[ItemA]}).
Hence, some of the nodes in the reachability graph must be pruned in order to restrict to valid states.

Coloured region in Figure~\ref{fig:pruned_rg} highlights the states of the feature-annotated reachability graph that are removed by the pruning strategies because of incompatible execution paths.

%% file: pruned_rg.tex
\scalebox{0.9}{
\begin{tikzpicture}[
  >=stealth,
  V/.style={draw, ellipse, inner sep=5pt, thick, node distance=10mm, font=\sffamily},
  every edge quotes/.append style={font=\sffamily, align=center,auto}
]
\begin{scope}[nodes=V]
\node (a) {Source(5)};

\node (b) [below left=of a]  {Source(3)ItemA(1)};
\node (c) [right=of b] {Source(2)ItemB(1)};

\node (d) [below left=of b]  {Source(1)ItemA(2)};
\node (e) [below=of b] {Source(3)Completed(1)};
\node (f) [below right=of b]  {ItemA(1)ItemB(1)};
\node (g) [right=of f]  {Source(2)Completed(1)};

\node (h) [below=of d] {Source(1)ItemA(1)Completed(1)};
\node (i) [right=of h] {ItemB(1)Completed(1)};
\node (j) [right=of i] {ItemA(1)Completed(1)};

\node (k) [below=of h] {Source(1)Completed(2)};
\node (l) [below=of i] {Completed(2)};
\end{scope}

\path[->]
(a) edge["startA/A"] (b)
(a) edge["startB/B"] (c)

(b) edge["startA/A"] (d)
(b) edge["endA/A"]   (e)
(b) edge["startB/B"] (f)

(c) edge["startA/A"] (f)
(c) edge["endB/B"]   (g)

(d) edge["endA/A"]   (h)

(e) edge["startA/A"] (h)

(f) edge["endA/A"] (i)
(f) edge["endB/B"]   (j)

(g) edge["startA/A"] (j)

(h) edge["endA/A"]   (k)

(i) edge["endB/B"]   (l)

(j) edge["endA/A"]    (l);

\begin{scope}[on background layer]
  \node[fill=blue!85!cyan!15,inner sep=0pt,rectangle,fit=(f) (i) (j) (l)] {};
\end{scope}
\end{tikzpicture}
}

%% file: tool_support.tex
\section{Tool Support}
\label{sec:tool_support}

We have developed \titan{} (\underline{T}ool for Petr\underline{i} net produc\underline{t} line \underline{an}alysis), which supports different techniques to analyse PNPL. It can be freely accessed at~\cite{Titan}. The following lines describe the tool architecture in detail.

\begin{figure}[h!]
    \centering
    \includegraphics[width=0.95\textwidth]{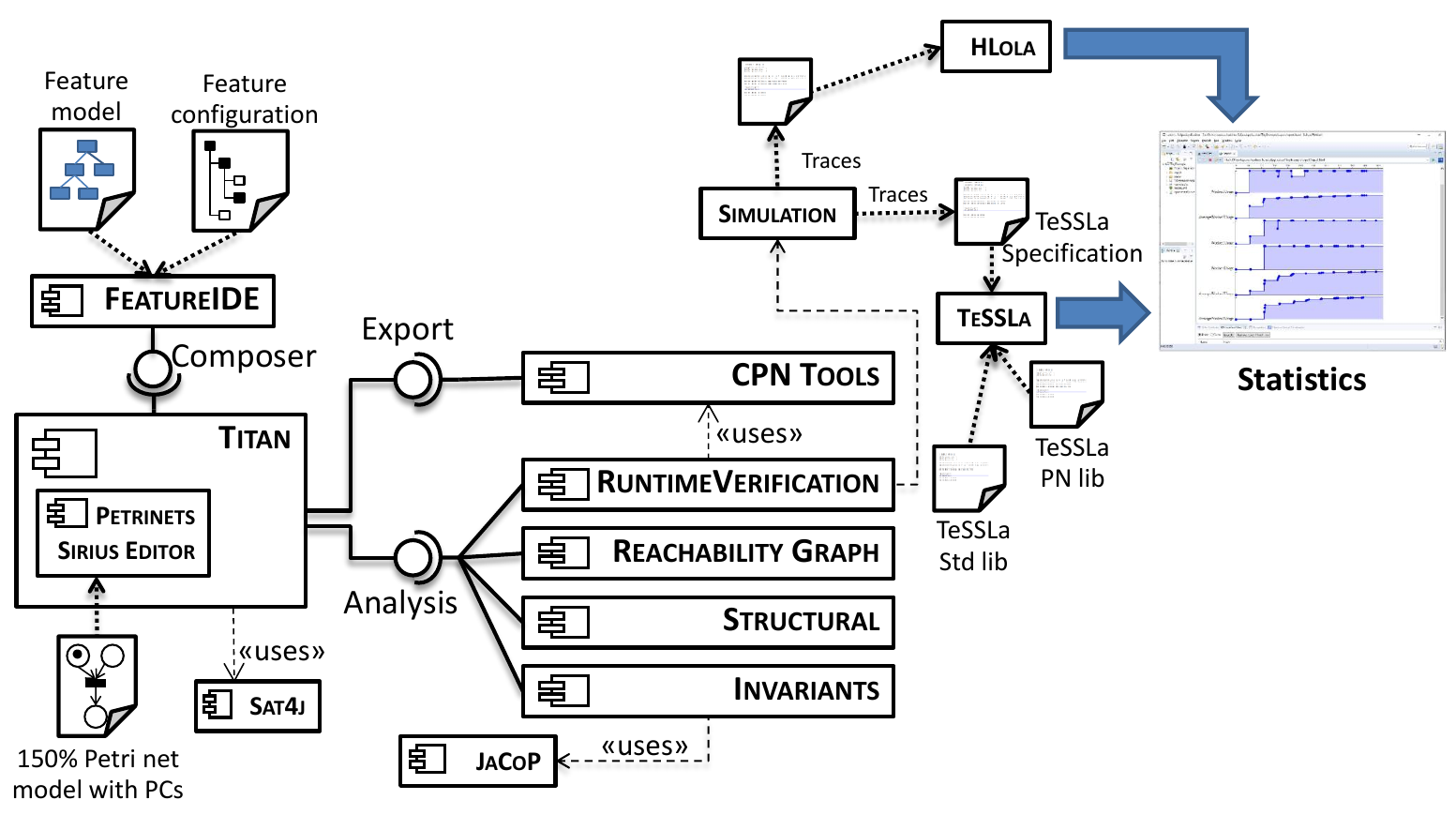}
	\caption{Architecture overview of \titan{}.}
	\label{fig:titan_architecture}
\end{figure}

\titan{} is a model-driven Eclipse plugin built in Java that leverages the Eclipse Modeling Framework (EMF)~\cite{EMF} for modelling and FeatureIDE~\cite{FeatureIDE} for defining feature models and configurations. It integrates a Sirius-based~\cite{Sirius} graphical editor to construct 150\% PNs and specify presence conditions (PCs). The extensible architecture supports additional analysis techniques and exporters via extension points. Figure~\ref{fig:titan_architecture} outlines the architecture.

\titan{} provides the lifted analysis of structural properties, such as marked graph, state machine, and (extended) free-choice using Sat4j as SAT solver, as described in~\cite{Gomez-MartinezLG21}. To analyse P- and T-invariants, \titan{} transforms the PNPL into a lifted matrix equation~\cite{Gomez-MartinezG23}. Thus, it is expressed as a Constraint Satisfaction Problem (CSP). CSPs are solved using constraint programming techniques, concretely it relies on JaCoP, a Java library for CP solver~\cite{Kuchcinski2013}. The generation of the reachability graph and state space exploration has also been implemented.
Moreover, \titan{} exports PNs (both 150\% net and any derivation net) to GreatSPN~\cite{Amparore15}, TimeNET~\cite{Zimmermann17} and WoPeD~\cite{FreytagS14}. It also supports the transformation of PNPL into CPN Tools~\cite{Gomez-MartinezG22}.


\titan{} currently computes reachability graphs and state spaces.
Recent additions allow the use of explicit time annotations in PNPL, resulting in Timed Petri Net Product Lines (TPNPL).
Time in TPNPLs aligns with the concept of time in Coloured Petri Nets \cite{Jensen1989,JensenKW07}, i.e., tokens carry discrete time annotations and time evolves according to a central clock.
\titan{} can perform runtime analysis of individual TPNPL products and compute user-specified quality metrics (e.g., performance, safety indicators) from execution events obtained by simulation traces. 
It relies on the TPNPL-to-CPN Tools transformation for dynamic analysis, enabling the selection of specific TPNPL products, execution of CPN Tools engines, and extraction of simulation reports.
To this end, \titan{} integrates two Stream Runtime Verification (SRV) frameworks (\tessla{}~\cite{ConventHLS0T18,KallwiesLSSTW22} and \hlola{}~\cite{CeresaGS20}) for runtime analysis.

\begin{figure}[h!]
    \centering
    \includegraphics[width=0.95\textwidth]{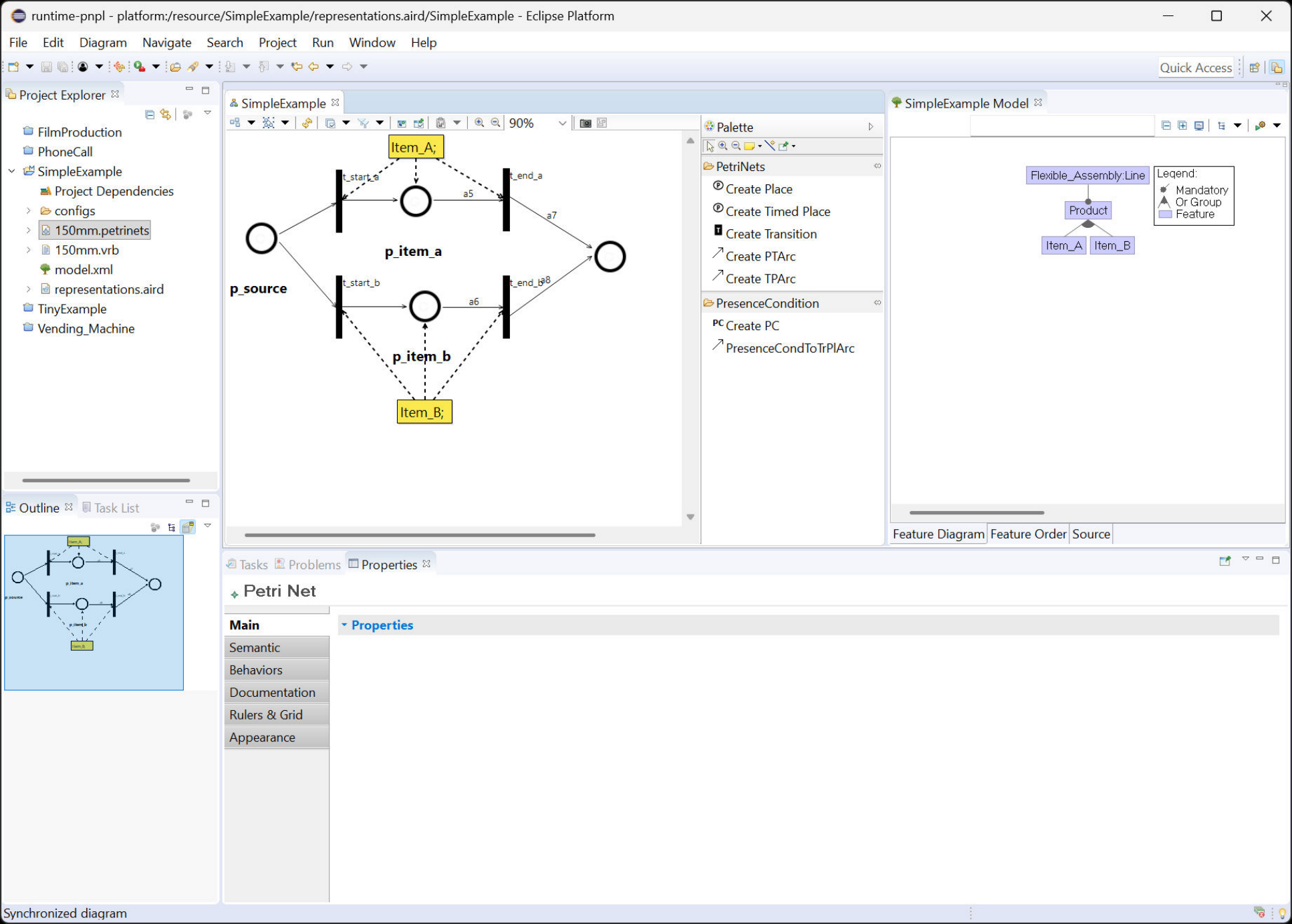}
	\caption{PNPL modelling of Figure~\ref{fig:running} within the \titan{} environment.}
    \label{fig:modelling}
\end{figure}

Figure~\ref{fig:modelling} presents a \titan{} snapshot with an example. The left panel shows the Eclipse explorer containing a FeatureIDE project with the PNPL from Figure~\ref{fig:running}. The central panel displays the 150\% net and its presence conditions, while the right panel shows the feature model.
Besides the PNPL in Figure~\ref{fig:running}, the \titan{} repository includes further examples, such as vending machines, flexible manufacturing cells, and the Haukeland University Hospital case study~\cite{PNSE_2024}.

%% file: related.tex
\section{Related work}
\label{sec:related}


The research presented in this work builds upon a rich foundation of work in Petri net theory and software product line (SPL) engineering, with a specific focus on extending these paradigms to handle variability for dynamic analysis. This section reviews the key contributions that derive feature-annotated reachability graphs (or, similarly, Featured Transition Systems - FTS$^+$) from Petri Net Product Lines (PNPLs) in order to explore system dynamics via state space exploration.

The verification of Petri nets and their extensions have been a subject of extensive research. \cite{Murata89} laid the groundwork for Petri net theory, while \cite{Petri62} provided detailed methods for reachability analysis. Software Product Lines (SPLs) are a software engineering methodology for developing a family of related systems by leveraging a common set of features and assets. There are several mechanisms to model variability for SPL and most of them can be classified into annotation-based and composition-based techniques \cite{Apel13}. Here are some works have added variability to Petri nets using SPL techniques.

Feature Petri Nets (FPNs) \cite{MuscheviciCP10} combine Petri nets with feature models by annotating net elements with propositional feature formulas, enabling feature-based enabling and disabling of behaviour. Dynamic Feature Petri Nets (DFPNs) \cite{MuscheviciPC16} extend this framework with reconfiguration transitions (e.g.\ \textsc{connect}/\textsc{disconnect}) that modify feature states at runtime, supporting the analysis of adaptive and context-aware systems.

Petri Net Product Lines (PNPLs) are designed to model a family of Petri nets representing an SPL, where variability is captured through a Feature Model (FM) and a labelling function $ \lambda $ that annotates transitions with feature conditions. The focus is on statically defining all possible product configurations and their behaviours.
The concept of PNPLs was pioneered by \cite{Gomez-MartinezLG21}, who proposed a framework to model variability in concurrent systems, building on earlier work in software product lines. This approach has been extended to support slicing and configuration analysis, but verification techniques remain underdeveloped.

In \cite{Gomez-MartinezG23}, the authors explored feature modelling in depth, providing tools for managing configuration spaces, yet their focus was on static analysis rather than dynamic behaviour. Regarding dynamic properties analysis of PNPLs, research on that is relatively limited.

Besides reachability graph, other methods support the analysis of dynamic properties of PN:  Linear Algebraic Techniques \cite{Desel96}, Simulation-Based Analysis \cite{Lu2019}, Model Checking \cite{Khomenko2003} and Transition Invariant Analysis \cite{Jensen1989}.
Among all these approaches, we chose extending reachability graph with feature annotations because the RG of a Petri net offers a complete representation of the state space: it systematically generates all possible states (markings) by firing enabled transitions from the initial marking $ M_0 $. This is essential for  the analysis of dynamic behaviour of PNPLs across all valid feature configurations, ensuring that properties such as reachability, deadlock, and liveness can be verified across the entire product line. 

Additionally, the RG analysis methods for Petri nets, developed by pioneers like \cite{Petri62} and \cite{Murata89}, include mature techniques such as state equations and coverability graphs. Although originally designed for classical Petri nets and do not extend to feature-annotated models, these methods are adapted for PNPLs through feature-extended RGs, leveraging their reliability to address variability-aware systems effectively.

%% file: conclusions.tex
\section{Conclusions}
\label{sec:conclusions}

This paper presents a formal framework and practical algorithms for constructing family-based reachability graphs of Petri Net Product Lines (PNPLs). Our primary contribution is a symbolic state representation that pairs net markings with presence conditions, complemented by family-preserving successor generation and conflict-aware pruning strategies that mitigate state-space explosion while preserving per-product semantics.

We prove the soundness and completeness of our construction with respect to traditional per-product analysis and analyse the main complexity trade-offs. An implementation within our PNPL modelling toolchain demonstrates significant efficiency gains over naive product enumeration, enabling scalable variant-aware verification, automated diagnostics, and design-space exploration for realistically sized product families.
We provide a preliminary version of the algorithm in the \titan{} tool, a plug-in for the Eclipse framework.

The proposed reachability-graph construction fills a critical gap in PNPL analysis and provides a solid basis for future work, including compositional verification, hierarchical net decomposition, richer temporal logic support, and tighter integration with existing model-checking and Petri net analysis toolchains.